%% file: templateArxiv.tex
\documentclass{article}

\usepackage{PRIMEarxiv}

\usepackage[utf8]{inputenc} 
\usepackage[T1]{fontenc}    
\usepackage{url}            
\usepackage{booktabs}       
\usepackage{natbib}
\usepackage{amsfonts}       
\usepackage{nicefrac}       
\usepackage{microtype}      
\usepackage{lipsum}
\usepackage{fancyhdr}       
\usepackage{graphicx}       
\graphicspath{{media/}}     

\usepackage{amsmath}

\usepackage{epstopdf}
\usepackage[title]{appendix}
\usepackage{flushend}

\usepackage{subfigure}
\usepackage[colorlinks=true, allcolors=blue]{hyperref}
\usepackage[table,xcdraw]{xcolor}
\usepackage{textcomp}
\usepackage[figuresright]{rotating}
\usepackage{tablefootnote}
\usepackage[justification=centering]{caption}
\usepackage{xcolor}

\title{Identifying Professional Photographers Through \\ Image Quality and Aesthetics in Flickr
\thanks{
\textcolor{red}{This is a pre-print version of the article.}} 
}

\author{
  Sofia Strukova, Rub\'en Gaspar Marco, Jos\'e A. Ruip\'erez-Valiente, F\'elix G\'omez M\'armol \\
  Department of Information and Communications Engineering \\
  University of Murcia \\
  Murcia (Spain)\\
  \texttt{\{strukovas, ruben.gasparm, jruiperez, felixgm\}@um.es} \\
}

\usepackage{amssymb}
\usepackage{subfigure}
\usepackage{color}
\usepackage{pifont}

\usepackage{epstopdf}
\usepackage{flushend}
\usepackage{graphicx}
\usepackage[table,xcdraw]{xcolor}
\usepackage{textcomp}
\usepackage[figuresright]{rotating}
\usepackage{tablefootnote}
\usepackage{multirow}

\usepackage{natbib}

\usepackage{color}
\usepackage{pifont}

\definecolor{ao}{rgb}{0.0, 0.5, 0.0}
\newcommand{\yestick}{{\color{ao}\ding{51}}}
\newcommand{\notick}{{\color{red}\ding{55}}}

\begin{document}
\maketitle

\begin{abstract}
In our generation, there is an undoubted rise in the use of social media and specifically photo and video sharing platforms. These sites have proved their ability to yield rich data sets through the users' interaction which can be used to perform a data-driven evaluation of capabilities. Nevertheless, this study reveals the lack of suitable data sets in photo and video sharing platforms and evaluation processes across them. In this way, our first contribution is the creation of one of the largest labelled data sets in Flickr with the multimodal data which has been open sourced as part of this contribution. Predicated on these data, we explored machine learning models and concluded that it is feasible to properly predict whether a user is a professional photographer or not based on self-reported occupation labels and several feature representations out of the user, photo and crowdsourced sets. We also examined the relationship between the aesthetics and technical quality of a picture and the social activity of that picture. Finally, we depicted which characteristics differentiate professional photographers from non-professionals. As far as we know, the results presented in this work represent an important novelty for the users' expertise identification which researchers from various domains can use for different applications.
\end{abstract}


\section{Introduction}
\label{sec:introduction}

Nowadays, we observe the emergence of a wide range of online technology-mediated portals. They have proved their ability to generate rich data sets through the users' interaction, which can be used to perform a data-driven evaluation of competencies and capabilities~\citep{strukova:survey}. Across them, there is a wide group of photo and video sharing platforms which are gaining indispensable popularity at the present time. This is explained by the fact that, in accordance with a comprehensive survey, users have five primary social and psychological motives to use one of the rising photo-sharing social networking services, which are social interaction, archiving, self-expression, escapism, and peeking~\citep{lee2015pictures}. More than that, images and photos are powerful tools based on their potential impact on people’s knowledge, attitudes, and perceptions regarding diverse topics~\citep{doi:10.1080/10584609.2019.1674979}. In this way, the mobile applications of most photo and video sharing platforms came onto the market at the right moment in the history of technology and made them the dominant image-sharing social media in the second decade of the 21st century~\citep{fung2020public}. 

Regardless of the high acceptance of photo and video sharing platforms throughout all segments of the world's population and their escalating use, there is no publicly available data set containing multiple data types and covering a considerable fraction of users from these platforms. Besides, despite the fact that there already exists ample evidence of vindicated methods used to measure the expertise of users across the group of sites that can be called content sharing and consumption~\citep{10.48550/arxiv.2204.04098}, there has not been much attention given to the in-depth exploration of photo and video sharing platforms that hold much potential to infer not only common metrics like popularity~\citep{ding2019social} but also a range of users' competencies or capabilities. One of the most valuable skills to detect and explore in this context is photography capability. As could be expected, photography skills are subjective and people often disagree with each other on the matter of taste. This is due to the fact that it is hard to conclude which photo is the best in terms of aesthetic and technical qualities. Since it is already not a trivial task for a person to identify technically sound and aesthetically attractive pictures, it is even more complicated for a machine to evaluate the quality of a picture explained by the fact that machines have to cope with noise in the picture represented by intensity levels, colour saturation, lighting, compression, artefacts, etc.~\citep{ding2019improved}. Also, machines do not have prior knowledge and struggle to understand some of the aspects of our world. As a good solution to the challenge of image preprocessing, Convolutional Neural Networks (CNN) trained with human-labelled data hold the potential to fill this gap~\citep{https://doi.org/10.48550/arxiv.2004.02168}.

Data generated on the photo and video sharing platforms hold the potential to be used in various contexts. Across them, we can highlight the possibility of the creation of pathways for learning about the user's behaviour, general traits of Web navigation and the ability to perform data-driven content analysis. More than that, this knowledge could be valuable for informal learning focused on acquiring new attainments or competencies~\citep{MEHRVARZ2021104184}. From another perspective, online content can yield violence in the user community, which is considered one of the most important problems of the 21st century~\citep{dikwatta2019violence}. In this way, data generated online hold the potential not only to infer valuable information about the users but also about vulnerabilities surrounding virtual life. Besides, the data set from any photo and video sharing platform with multimodal data would be able to infer the photography capabilities of users. This will open an opportunity to automatically detect good photographers on the Web and offer personalised aesthetic-based photo recommendations.

In this work, we examine several photo and video sharing platforms and the existing studies focusing on analysing data available across them. Based on these grounds and the encountered gaps, our first step was to create one of the largest data sets available from the Flickr platform~\citep{gasparmarco2022dib}. We collected data from 27,538 users who uploaded photos to Flickr in December 2021, specifically those who specified their occupation. Additionally, we enriched the data set with features resulting from the automated analysis of the photos and their comments including three Image Quality Assessment scores representing aesthetic and technical aspects of the photos. Also, we labelled the data to indicate whether the user is a professional photographer. We are releasing the data set as part of this papers' contribution. Thus, it is open sourced and is available in the following URL:~\citep{dataset_flickr}. Next, we propose our method to infer if a user is a professional photographer or not based on self-reported occupation labels, which is a novel contribution to the literature. Finally, to the best of our knowledge, this is the first time that characterisation of professional and non-professional users is presented in any photo and video platform. 

Accordingly, the first objective of the paper at hand was to create a data set focused on the Flickr photo and video sharing platform with multimodal data including crowdsourced, user and photo features that would allow to answer the following Research Questions (RQs) that we state next:

\begin{itemize}
    
    \item RQ1. What model is better to infer if a user is a professional photographer? One based on photo features including aesthetics and technical quality scores, one based on the social network activity of the photographer, or one based on crowdsourced features that represent the interaction of other users with the photo? 
    
    \item RQ2. What is the relationship between the aesthetics, the technical quality and the social activity of a given picture?
    
    \item RQ3. What characteristics differentiate professional photographers from non-professionals?
    
\end{itemize}

The remainder of this paper is structured as follows. In Section~\ref{sec:background}, we focus on the background of our study uncovering the subject of photo and video sharing platforms. In Section~\ref{sec:methodology}, we present our research methodology. We expand this section by selecting the photo and video sharing platform and explaining the data collection process. Next, we depict the final data collection and describe machine learning (ML) algorithms to identify professional photographers. Our findings are outlined in Section~\ref{sec:results}, while we extend the results in Section~\ref{sec:discussion}. Finally, we draw our conclusions and future research directions in Section~\ref{sec:conclusion}.

\section{Background}
\label{sec:background}

\subsection{Photo \& Video Sharing Platforms}
\label{subsec:photo_video_platforms}

The main goal of photo and video sharing platforms is to allow their users to share various multimedia content, including photos and videos. Some of the platforms have built-in editing filters and organisation by hashtags and geographical tagging. Most of these sites also include a social networking service permitting users to connect with each other through comments or messages, browse other users' content, share and receive feedback. In this way, some material can be shared publicly or with pre-approved followers. In Table~\ref{tab:platforms_comparison}, we present a comparison of the leading photo and video sharing platforms, namely, Flickr, 500px\footnote{\url{https://500px.com/}}, Instagram\footnote{\url{https://www.instagram.com/}}, 1x.com\footnote{\url{https://1x.com/}}, SmugMug\footnote{\url{https://www.smugmug.com/}} and Pinterest\footnote{\url{https://www.pinterest.com}}, across several characteristics.

\input{tables/platforms_comparison}

There were three fundamental points of comparison for our research: the number of monthly users, the access to an Application Programming Interface (API), and the ability to write comments. We could not find the up-to-the-minute number of active users per month in 500px, 1x.com and SmugMug, while among others, the most visited portal is Instagram with its 2,000 million users per month, followed by Pinterest and Flickr with their 430 and 90 million users, respectively. Finally, most of the portals that we explored offer an API, except 1x.com and 500px that shut down its API access in 2018.

Flickr was a pioneer in online photo sharing and nowadays is one of the leading photo-sharing platforms worldwide which attracts extensive research attention~\citep{hopken2020flickr}. Its users include diverse profiles of both professional and amateur photographers who want to share their portfolios. In 2018, it was acquired by SmugMug, a paid photo-sharing service. Similarly, SmugMug is characterised as a premium online photo and video sharing service business which currently has material uploaded by amateur and professional photographers around the world~\citep{erturk2016multimedia}. 500px and 1x.com are also more suitable for serious cameramen and they offer an image-focused design. On the contrary, Instagram is a social photo-sharing service launched in 2010 as an iPhone application fitting for non-professional users. Its users can take and manipulate photographs by adding filters and frames that enhanced the users’ experience. They can also share them online where other users can react by means of comments and ``likes". Instagram is bringing an opportunity to communicate experiences through both choice of photo subject and ways to manipulate and present them~\citep{10.1145/2470654.2466243}. Lastly, Pinterest was launched in 2010 as a Web site where users can save an image (known as a ``pin") that they upload or find on a Web page onto a collection of these pins. A more detailed description of these platforms can be found in~\citep{10.4018/978-1-6684-3996-8.ch005}.

\subsection{Related work}
\label{subsec:related_work}

There exist many studies disclosing the potential of Web portals to yield a significant amount of data, which can allow the detection of potential experts. On the whole, the expertise finding is focused on detecting topical authority in a selected topic in forums and question and answer websites (e.g., Reddit~\citep{10.48550/arxiv.2204.04098, lim2017estimating} or Quora~\citep{patil2016detecting}). In contrast, most of the research in this domain is centred on proficiency in different programming languages, libraries or tools across portals highly related to the field of computer science such as GitHub~\citep{saxena2017know} and StackOverflow~\citep{constantinou2016identifying}. However, there is not much work done on discovering artistic skills which are crucial to look at things from different perspectives and to remain competitive globally~\citep{doi:10.1080/23311983.2022.2043997}. We also did not find any study aiming to identify professional photographers through image quality, aesthetics or any other photo-related features, thus both our novel data set~\citep{gasparmarco2022dib} and our research in this study significantly contributes to the literature.

From another perspective, the vast majority of the studies are making use of single-mode data sources. For example, Kantharaju et al. utilised clickstream data to trace player knowledge in educational games~\citep{kantharaju2019tracing} and Pal et al. extracted textual data represented in questions and answers of users of a question and answer portal~\citep{pal2011early}. On the contrary, very few researchers decided to employ multimodal data sources. One of the examples of such an approach is~\citep{van2015early} demonstrating the use of textual, behavioural and time-aware features in StackOverflow. The results of this work proved the utility of adding behavioural and time-aware features to the baseline method with an accuracy improvement for early detection of expertise. Even though there is a clear trend in using multimodal data, we did not find previous studies that operated various types of data in photo and video sharing platforms.

Also, we saw a heightened interest towards photo and video sharing platforms which could be able to reflect important information about users. A few studies are revealing that rich data sets from these portals could be used to explicitly or implicitly perform a data-driven evaluation of diverse capabilities. For example, Pal et. al. presented a novel approach to finding topical authorities in Instagram~\citep{10.1145/2872427.2883078}. Their method is based on the self-described interests of the follower base of popular accounts. Similarly, Purba et al. carried out an analysis of popularity trends and predictions on Instagram, using a set of features acquired from users’ metadata, posts, hashtags, image assessment, and history of actions~\citep{purba2021instagram}. In the analysis of popularity trends, engagement grade is used in comparison to respect the lower engagement rate of users with a higher number of followers. It was found that image quality, posting time, and type of image highly impact engagement rate. However, neither of these studies of Instagram focused on photography capabilities as we do.

Finally, despite the enhancing relevance of photo and video sharing platforms and research across them, there are no publicly available data sets that could be used for the exploration of users' personality traits or capabilities. This is an important gap existing in the current domain.

\section{Methodology}
\label{sec:methodology}

In this section, we describe the methodology process of building a supervised learning model for the final goal of professional photographers' identification. First, we explain the photo and video sharing platform selection followed by the description of its API service. Next, we give details on the feature engineering process. Then, we describe the final data collection and the ground truth. Finally, we explain the ML models that we chose for the stated goal and evaluation metrics to estimate their performance.

\subsection{Methodology overview}

\begin{figure*}[!ht]
\includegraphics[width=1\textwidth]{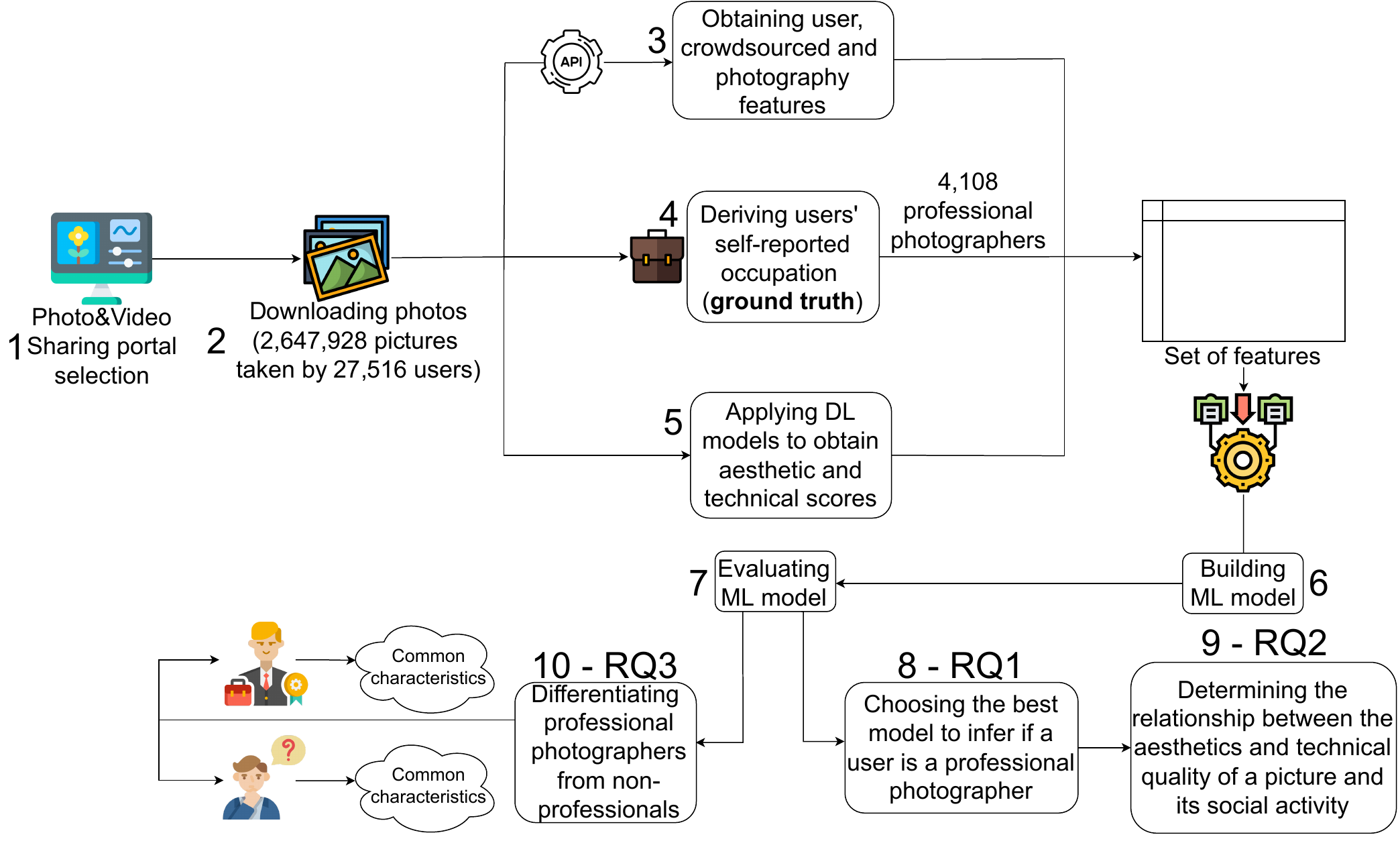}
\centering
\caption{Overview of the methodology to identify professional photographers in Flickr}
\label{fig:methodology}
\end{figure*}

To answer the RQs stated at the beginning of our study, we pursued the methodology process presented in Figure~\ref{fig:methodology}. 

In the first step, we selected the photo and video sharing platform based on various metrics presented in Table~\ref{tab:platforms_comparison}. In the second step, we downloaded the photos from the selected site. In the third, fourth and fifth steps, we obtained the user, crowdsourced and photography features and ground truth in order to build the ML model in the sixth step. In the eighth step, we chose the best model to infer if a user is a professional photographer or not based on self-reported occupation labels. Next, in the ninth step, we explored the relationship between the aesthetics and technical quality of a picture and its social activity. Finally, in the tenth step, we found the common characteristics of professional photographers and non-professionals.

\subsection{Photo and video sharing platform selection}

Based on Table~\ref{tab:platforms_comparison} presented in the previous section, we can conclude that the most active photo and video sharing platforms are Instagram, Pinterest and Flickr, being visited by 2,000, 430 and 60 million active users every month, respectively. Flickr has a PRO service where a user can get unlimited storage, making it one of the cheapest hosting sites around. To keep Pinterest running smoothly, the users can create up to 200,000 pins and 2,000 boards which is a collection where users save specific pins. In contrast, Instagram allows its users to upload an unlimited quantity of photos. Moreover, as discussed earlier, these three platforms offer API services that can be helpful while acquiring the needed data. 

Although Instagram has the huge privilege of having many registered users and the facility of uploading an unlimited number of photos, for our study we see it as a disadvantage because it could be hard to find those users whose behaviour on the website would correspond with the profile of professional photographers. Moreover, it does not allow uploading original-sized photos. On the other hand, Pinterest is not focused on pictures taken by users themselves but rather on drawings, paintings or artworks created on a computer. Therefore, we will focus on a photo and video sharing platform Flickr as a proxy for the photography skills of users.

Flickr also differs from Instagram by providing online communities and other groups on numerous social media and other platforms to improve customer relationships. Groups are a place to share ideas and photos with other like-minded members. Some group administrators first have to approve the users’ request to join. Flickr offers its users to create profiles with personal information, albums/photosets which are helpful to organise their photos and galleries/collections to which they can add other users’ media. Flickr is also geared toward beginners and enables them to edit the photos directly on the platform, such as adjusting brightness and contrast and applying various filters. There is also a concept of photostream which is a collection of media files that solely belongs to a user (public -- others can visit the profile and see what the user uploaded, private -- only the user and the list of permitted users will be able to view the content). All users have a list of their favourite photos. There is also an ability to connect with other users. As a social photo-sharing site, Flickr allows users to maintain a list of contacts. From the perspective of a registered user of Flickr, there are five categories of people on Flickr: the user, the user’s family, the user’s friends, the user’s contacts who are neither family nor friends, and everyone else~\citep{yee2008pro}. Statistics for a free account show the total number of views, favourites, and comments it has. From another point of view, users can use tags to categorise and search for photos. There are several ways to tag pictures, either one at a time or in batches. Flickr lets users add up to 75 tags to each picture including the geotagging feature. Finally, every user can set a license representing the copyright permission for a given picture.

\subsection{Flickr API -- data collection}

Flickr provides an API service which facilitates significantly the process of data collection. Primarily, we decided to download the data of only those users who had filled the occupation field in their profile. This selection is explained by the fact that we wanted to avoid bias derived from assuming their profession. Intending to obtain a representative and comprehensive sample of the platform's active users, we singled out those users who were sufficiently active during the month of December 2021. We searched for all the photos of the month of December discarding screenshots and videos. There were 225,590 users who uploaded photos in December 2021.

For the user selection, we discarded those users whose number of photos uploaded in December 2021 was equal to or greater than 20\% of their total activity, in order to filter out those users without a minimum activity on the platform. We also filtered out the 5\% of users from both ends of the distribution of total photos uploaded to avoid outliers. As a result, the final number of users we selected is 151,468.

For the time limits reason, it was impractical to aim to extract the data from all the photos from all the users. Finally, we downloaded all pictures of the selection of 27,538 users. The complete process of downloading data with Flickr API is thoroughly explained in~\citep{gasparmarco2022dib} which also describes the full data collection process.

\subsection{Feature engineering for the ML model}

\subsubsection{Deep learning models}
\label{subsec:dl_models}

Child states that the photographer has to pre-visualise, pre-produce and create an environment using not only selected equipment, subject matter, props, and far more importantly, light~\citep{child2013studio}. Image quality can be affected by the noise, the blur and the used technical requirements and equipment. From another perspective, the aesthetics of the photo depend on the colours' balance (their compatibility and what feelings they evoke), contrast (variance between light and dark), lighting in general, camera to subject diagram, camera angle and height, meter readings of light ratios, composition, subject choice and symmetry. 

Despite the fact that evaluating these points might be hard for an ordinary user, some models perform well in this regard. After exploring several surveys including a comprehensive performance evaluation of image quality assessment algorithms~\citep{8847307}, we selected two algorithms based on their high performance and the ability to rebuild them. Firstly, Neural Image Assessment (NIMA) -- a deep CNN that is trained to predict which images a typical user would rate as looking good (technically) or attractive (aesthetically)~\citep{talebi2018nima}. Other models classify images as low/high scores while the NIMA model produces a distribution of ratings for any given image -- on a scale of 1 to 10, NIMA assigns likelihoods to each of the possible scores. Various functions of the NIMA vector score (such as the mean) can then be used to rank photos aesthetically. The authors replaced the last layer of the baseline CNN with a fully-connected layer with 10 neurons followed by soft-max activations. Baseline CNN weights are initialised by training, and then an end-to-end training on quality assessment is performed.

Secondly, Photo Aesthetics Ranking Network with Attributes and Content Adaptation proposes to train a deep convolutional neural network to rank photo aesthetics in which the relative ranking of photo aesthetics is directly modelled in the loss function~\citep{kong2016photo}. This model incorporates joint learning of meaningful photographic attributes and image content information which can help regularise the complicated photo aesthetics rating problem. This model returns ratings for any given image -- on a scale of 0 to 1.

\subsubsection{Comments preprocessing}
\label{subsec:comments_preprocessing}

To analyse the comments, we followed several data preprocessing steps for comments. First, we changed all the comments to lowercase. Next, we replaced emojis in comments with their description codes with the use of the Python Demoji library~\footnote{\url{https://pypi.org/project/demoji/}}. Moreover, we cleaned the comments from hyperlinks and non-alphanumeric text. Finally, we removed empty comments (including those that consisted only of stop words). For every comment, we computed features further explained in Section~\ref{subsec:final_data_collection}. 



\subsubsection{Description of the final data collection}
\label{subsec:final_data_collection}

The final data set used for further investigation consists of 2,647,927 pictures of 27,538 users. Each picture was downloaded in a size such that the smallest side of the image measures more than 230 pixels because the NIMA model takes images of 224x224 pixels size as input and Photo Aesthetics Ranking Network with Attributes and Content Adaptation works with 227x227 pixels images.

We have grouped the features that we obtained into three families -- photography, crowdsourced and user-author. The photography feature set includes the following features:

\begin{enumerate}
    
    \item Publication date -- Number of days since the photo uploaded to Flickr.
    
    \item Update date -- Number of days since the last update of the photo metadata (visits, favourites, comments, etc.).
    
    \item Groups number -- Number of groups in which the photo has been posted.
    
    \item NIMA technical score -- Technical score from NIMA model implementation.
    
    \item NIMA aesthetic score -- Aesthetic score from NIMA model implementation.
    
    \item Kong score -- Aesthetic score from Photo Aesthetics Ranking Network with Attributes and Content Adaptation implementation.
    
\end{enumerate}

Crowdsourced features. These involve information or opinions from a group of people who submit their views via the Flickr site.

\begin{enumerate}

    \item Comments number -- Number of comments written on the photo page.
    
    \item Views number -- Number of views the photo got.
    
    \item Favourites number -- Number of users who added the photo to the list of their favourites.
    
    \item Average polarity of the comments -- computed with TextBlob. TextBlob a Python library for processing textual data~\citep{loria2018textblob}.
    
    \item Average subjectivity of the comments -- Average number of subjective words in posted answers computed with TextBlob.
    
    \item Average readability of the comments, including two metrics indicating how difficult a passage in English is to understand, such as the number of difficult words and reading time.
    
    \item Average entropy of the answer -- A statistical parameter that measures how much information is produced on average for each letter of a text in a language.
    
    \item Average comment length -- Character count of the comment.

\end{enumerate}

The user feature set includes the following features:
    
\begin{enumerate}
    
    \item Photos number -- Total number of photos uploaded by the user to the platform.
    
    \item Join date -- Number of days since the user became the member of the forum.
    
    \item Following number -- Number of the users followed by the user.
    
    \item Groups number -- Number of groups to which the user belongs.
    
     \item Flickr PRO -- The indication if the user has the paid membership Flickr PRO\footnote{\url{https://www.flickr.com/account/upgrade/pro/}}. Flickr PRO provides advanced statistics on photos and videos of the user. Also, it allows ad-free browsing on Flickr for the PRO user and their visitors. Moreover, it permits unlimited uploads at full resolution and easily backup. Finally, the user can establish detailed privacy settings for every photo.
    
\end{enumerate}

Next, we aggregated crowdsourced and photo features by every user. As a result, for every user, there is a representation of every feature in terms of a minimum, a maximum and an average value.

Finally, for a better understanding, as social activity features to answer the RQ2, we consider a minimum, a maximum and an average values of the following variables -- the number of comments and the number of favourites (how many people added a picture to their list of favourites).

\subsubsection{Ground truth}
\label{subsec:ground_truth}

To collect the ground truth values, we firstly obtained the occupation self-indicated by the user. Based on it, we detected if the occupation is related to photography. It was computed with the use of regular expressions in several languages that use the Latin alphabet. The regular expression includes the following terms -- ``fot", ``phot", ``valokuv", ``zdj\c{e}cie", ``dealbh", ``bild", ``grianghraf", ``nuotrauk", ``pictur", ``myndin", ``billed", ``lj\'osmyndari", ``ritratt". Accordingly, as a ground truth value, we will consider those users who have the photography-related occupation. This being the case, there are 4,108 users ($\approx$15\%) fulfilling this criteria.


\subsection{ML model to identify professional photographers}
\label{subsec:ml_models}

Following the scope of our study, we will compare the performance of both interpretable and non-interpretable classification techniques over the features mentioned in the previous section to fulfil the goal stated in Section~\ref{sec:introduction}. We selected the two most interpretable classification techniques including a probabilistic and a logit models -- Gaussian Naïve Bayes and Logistic Regression (LR) and pit them against non-interpretable techniques including a bagging and a boosting models -- Random Forest (RF) and Gradient Boosting Classifier. We believe that our choice of algorithms covers a wide spectrum of attribute-based learning approaches. Hence, we restrict our case study to the use of these algorithms with the final goal of selecting the best model out of a set of classifiers with various feature representations.

We have trained our model applying a 10-fold cross-validation. Given the significant data imbalance, we have trained the model to maximise the quality metric AUC, which takes into account the data imbalance. Moreover, we also report F1-score and the accuracy of the model.

\section{Results}
\label{sec:results}

\subsection{RQ1. Professional and non-professional photographers}

From Table~\ref{tab:results_models}, we can observe the fact that all the models -- interpretable which include Gaussian Naïve Bayes and LR and non-interpretable ones represented by RF and Gradient Boosting Classifier -- perform in a similar way. The results show that, in general, most competing algorithms were fairly accurate for our data set. The surprising observation is that the accuracy score varies from 0.85 and reaches 0.92 in most tests of models and feature sets. This can be explained by the fact that accuracy is a simple evaluation measure for binary classification and it is more suitable for matters when the data are perfectly balanced. This is not the case in our study and it was explained previously in Section~\ref{subsec:ground_truth}. Consequently, we computed AUC and F1 scores which help us to observe the precision and the recall providing more insight into the differences between classifiers.

The comprehensive comparison revealed that RF demonstrates the best performance when using the user and photo features reaching an accuracy of 0.92, an AUC score of 0.73 and an F1 score of 0.89. Similarly, with RF, Gradient Boosting Classifier was almost as successful as the approach with the best capacity. It showed just slightly worst results if we compare the AUC and F1 measures of each model in every set of features. Nevertheless, its prediction with the set including user features showed sufficiently good results indicating an accuracy of 0.92, an AUC score of 0.72 and an F1 score of 0.89. In this way, non-interpretable algorithms exhibited superior evaluation performance compared with the rest of the contenders.

\input{tables/results_models}

Regarding the comparison of interpretable models, among Gaussian Naïve Bayes and LR, the best-performing model is LR with the photo features indicating an accuracy of 0.92, an AUC score of 0.68 and an F1 score of 0.88. From Table~\ref{tab:results_models}, we note that the performance of Gaussian Naïve Bayes is less competitive compared to LR. However, even though these algorithms achieve the lowest AUC score, they reach comparably high accuracy and F1 score. We also can observe that combining sets of features does not always mean a clear increase in the evaluation ability than using each set of features alone.

\subsection{RQ2. The aesthetics and technical quality and the social activity of photos}

\begin{figure*}[!ht]
\includegraphics[width=1\textwidth]{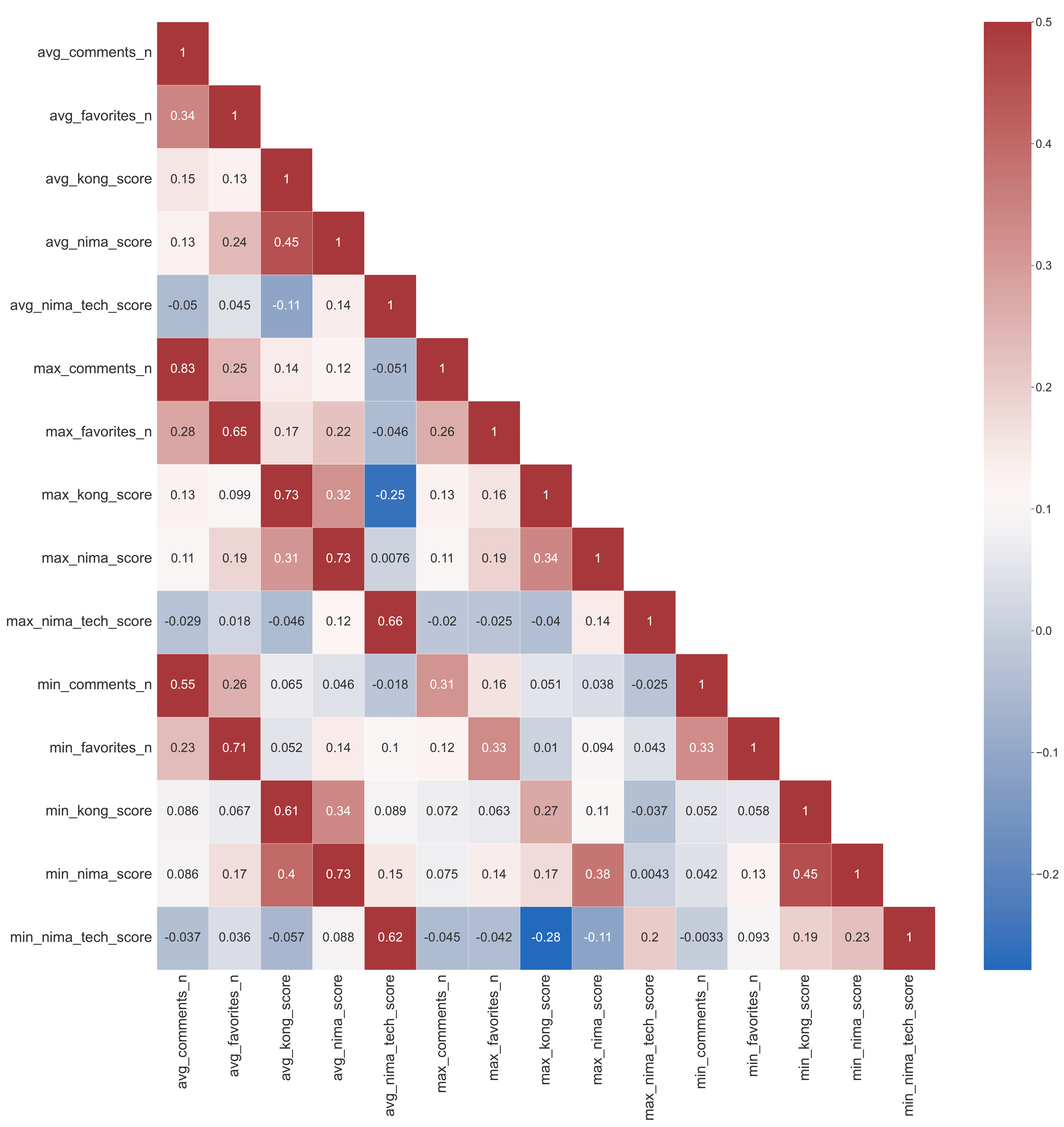}
\centering
\caption{Correlation matrix between the social activity features and NIMA technical score, NIMA aesthetic score and Kong score}
\label{fig:correlation_matrix}
\end{figure*}

To answer the question regarding the relationship between the aesthetics and technical quality of a picture and the social activity of that picture, we first explored the correlation matrix of the social activity features and NIMA technical score, NIMA aesthetic score and Kong score represented in Figure~\ref{fig:correlation_matrix}. As it is plain to observe, the technical and aesthetic scores are highly correlated between themselves. As a case in point, the correlation between the average of the Kong scores and the average of the NIMA aesthetic scores is 0.45 which indicates that they are strongly positively correlated. As depicted in the figure, there are many variables that are relatively highly correlated with their different representations (a minimum, a maximum and an average). Variables that represent different concepts are not positively correlated with each other, with the exception of aesthetic and technical scores.

\input{tables/rq2_results}

Then, we examined the performance of the best-performing model selected in the previous section -- RF separately on the social activity feature set and the features related to the aesthetics and technical quality of pictures. The results reported in Table~\ref{tab:rq2} indicate that the algorithm has a better predictive power with the photo features that include NIMA technical score, NIMA aesthetic score and Kong score reaching an accuracy of 0.92, an AUC score of 0.67 and an F1 score of 0.88. On the other hand, with the social activity set of features, Gradient Boosting Classifier shows a lower AUC score equal to 0.6 but the same F1 score equal to 0.88. This means that despite the subjectivity of art, the aesthetic and technical scores computed by CNN models are reliable.

\subsection{RQ3. Common characteristics of professional and non-professionals}

\begin{figure*}[!ht]
\includegraphics[width=1\textwidth]{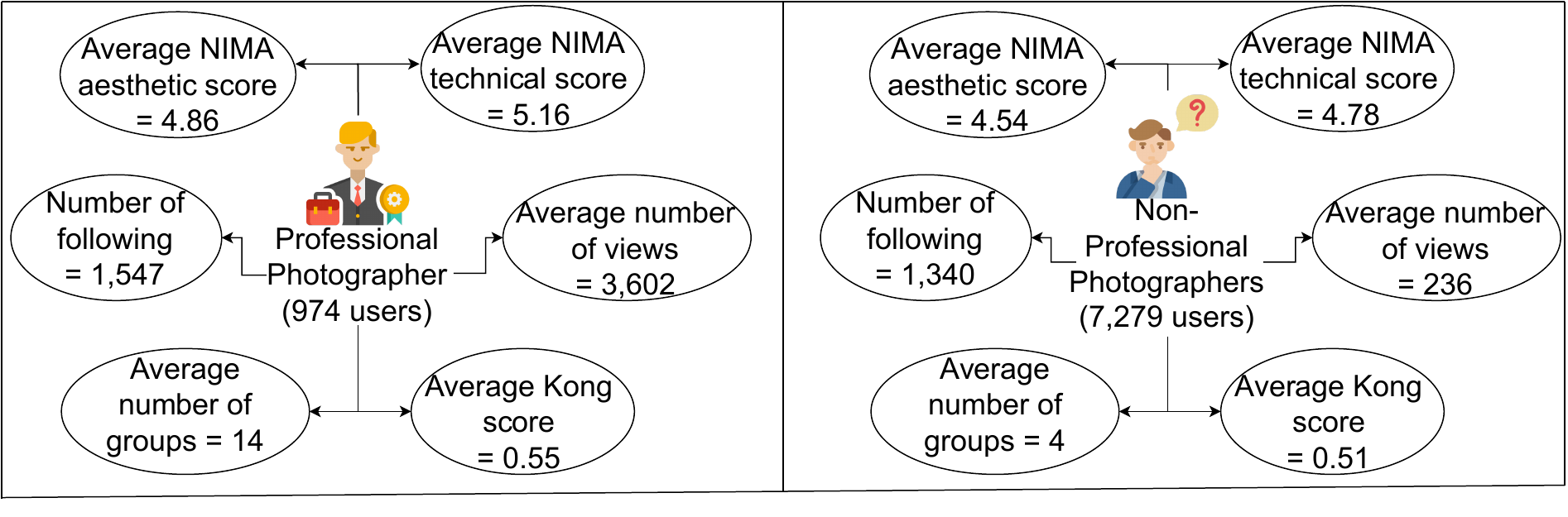}
\centering
\caption{Aggregation of professional photographers and non-professional users with their most common characteristics}
\label{fig:user_aggregation}
\end{figure*}

To answer this RQ, we aggregate the results of the prediction of the RF best performing model with user and photo features. We computed average metrics per professional photographer and non-professionals for all user and photo features that were explained in Section~\ref{subsec:final_data_collection}. The aggregation of these two types of users predicted by RF with their most common characteristics and differences is shown in Figure~\ref{fig:user_aggregation}. The model identified 974 users as professional photographers ($\approx$12\%) and 7,279 users as non-professionals. It is noteworthy to mention that there are many more non-professional users than other types of users. It is consistent with the distribution of ground truth presented in Section~\ref{subsec:ground_truth} where we explained the class imbalance context.

In this Figure~\ref{fig:user_aggregation}, we depict these two types of users and the metrics which differentiate them. We performed a multivariate analysis of variance (MANOVA) to ascertain that the differences between these types of users are statistically significant. This fact was confirmed by obtaining an $F$-value = 3,268 and $p$-value $ \ll $ 0.0. Thus, we can confirm that the two types of users have statistically significant different characteristics. Also, we conducted an analysis of variance (ANOVA) for each individual feature to see which of them are statistically different which are represented in Figure~\ref{fig:user_aggregation}. The outcomes of average photography technical and aesthetic scores show that photos of professional photographers got a higher NIMA aesthetic score (4.86 versus 4.54), NIMA technical score (5.16 versus 4.78) and Kong score (0.55 versus 0.51). Moreover, photos of professional photographers tend to be visited more often getting an average number of views equal to 3,602 while photos of non-professional users get an average of 236 views. Besides, pictures of professional users differ from another type of users by the average number of groups where they are published -- 14 versus 4. Finally, the number of users followed by professional photographers is fairly higher -- 1,547 opposite 1,340 in the case of non-professional users.

\section{Discussion}
\label{sec:discussion}

In this section, we first discuss the obtained results. Then, we talk about the potential application of our study in real scenarios. Finally, we raise the limitations of our work.

\subsection{Obtained results}

To summarise, the experiments conducted on the data set extracted from Flickr suggest that all competing interpretable and non-interpretable algorithms (Gaussian Naïve Bayes and LR, RF and Gradient Boosting Classifier) provide meaningful results. It is worth mentioning that studies examined in Section\ref{subsec:related_work} applied models for the task of expert finding in technical fields or concise areas which showed better results. This can be explained by the fact that artistic skills and specifically photography skills are ill-defined and there are no established methods and features to determine and measure them. Also, based on the results presented in Table~\ref{tab:results_models}, we can notice that some feature sets are more informational than others. For example, the results of using photo and user features are proving that these denominate more meticulously professional users. This indicates that the Flickr data of users and photos can be used to identify professional photographers and non-professional users based on self-reported occupation labels. Other researchers can use our findings as a base to find more powerful models in order to strengthen the detection of experts in the photography field.

After recognising the satisfactory performance of the above-mentioned algorithms, we focused on RQ2 questioning the relationship between the aesthetics and technical quality of a picture and the social activity of that picture. The fact that we did not see much correlation between social activity features and technical and aesthetic scores of the photos was not unpredictable. It can be explained by the fact that many existing studies in the literature on image aesthetic assessment are based on the data sets like A Large-Scale Database for Aesthetic Visual Analysis~\citep{6247954} or Tampere Image Database~\citep{PONOMARENKO201557}. These data sets were annotated with semantic and aesthetic labels and rated by users unidentifiable to researchers. In this way, it is not clear that these annotators align with the photography enthusiast community. Besides, aesthetic beauty is subjective based on the fact that the perception of the beauty of the same picture can be different. Moreover, due to the social network features of the photo and video sharing portal, the behaviour of users might prevail over aesthetics.

Ultimately, to answer the RQ3 in relation to the characteristics that differentiate professional photographers from non-professionals, we aggregated the results of the prediction of RF by every user. Based on the statistically significant features, we noticed that users differ by the average NIMA aesthetic score, the average NIMA technical score, the average Kong score, the average number of groups, the average number of views and the number of the following users. All these features are reasonably higher for professional photographers. The fact that photos from non-professional users are less visited can be explained by two other variables -- the average number of followers and groups. The number of followers the users by default explains the number of clicks that their pictures could get. On the other hand, to achieve that a picture is published in a group, the user should conduct some activity by uploading it there. Besides, some groups require administration approval for the picture to be in the group. We can conclude that definitely this is correlated with the technical and aesthetic scores which are higher for professional photographers.

\subsection{Application in real scenarios}

The results presented in this paper can find workability in the task of assessing the photography quality of users. The automatic detection of professional photographers can be used in order to build more reliable photo and video sharing platforms by establishing high standards of skills for users.

Moreover, our findings can be applied in different contexts apart from the stated identification of professional photographers. Through computer vision, we can detect inappropriate content. This is a relevant issue nowadays since the proliferation of social media enables people to express their opinions widely online leading to the emergence of conflict and hate. The lack of a universal hate classifier generalising various training sets and contexts was addressed by~\citep{salminen2020developing}. The authors developed a cross-platform online hate classifier which performs well for detecting hateful comments across multiple social media platforms including YouTube, Reddit, Wikipedia and Twitter. However, the data sets used for this study mainly include manually-labelled comments from these sites. We believe that this work can contribute significantly to improving the coverage of the existing platform. More than that, in these latter days, even the seemingly harmless meme can become a multimodal type of hate speech considered as a direct attack on people based on ethnicity, religious affiliation, gender, etc.~\citep{velioglu2020detecting}. 

More than that, our results can serve as a base for creating a platform which could offer personalised aesthetic-based photo recommendations. This tool is already implemented in several portals such as the Netflix recommender system~\cite{10.1145/2843948} and there is a need of extrapolating it to other platforms. It can help photography websites better serve the needs of non-professionals and professional photographers~\cite{zhou2018aesthetic}. Content-based image search does not fully satisfy the needs of such users since they are usually not interested in content alone. Instead, they are often looking for photos with certain photographic aesthetics, which may include monochromaticity, light contrast, and style.

Another important topic that can be addressed with the help of this study is privacy issues. We can detect sensitive places and photographs violating the community terms and conditions. On the one hand, most websites nowadays are taking measures against spam messages and inappropriate content. However, every day malefactors are inventing new ways of overcoming them. Also, not all the systems can detect photographs of which places can make it vulnerable.


\subsection{Limitations}

We should like to discuss the research gaps that arise today within the topic of the identification of professional photographers on Flickr.

It is important to mention that our results prove the potential of ML models to be used in several domains. However, the expected foundation of image quality and aesthetics was not proved for professional identification. We noticed a certain level of correlation assumed from the MANOVA test but not enough to assure that pictures of professionals have higher scores according to the models described in Section~\ref{subsec:dl_models}. 

More than that, the ground truth for this study is based on the self-proclaimed occupation of the users. We believe that there might be more characteristics of professional users that could be used as the base of the ML model.

Finally, it would be useful to repeat the study with other portals explored in Section~\ref{subsec:photo_video_platforms}. Flickr, being a large photo and video sharing platform, could be not representative enough regarding the amount of not professional users.

\section{Conclusions and Future Work}
\label{sec:conclusion}

This work aimed to fill the gap of the lack of any open data set on photo and video sharing platforms. We provided a significant contribution to the literature by collecting one of the largest labelled data sets on Flickr with multimodal data including crowdsourced, user and photo features. From 225,590 users who uploaded photos in December 2021, we filtered out users without a minimum activity on the platform and the 5\% of users from both ends of the distribution of total photos uploaded to avoid outliers. As a result, the final number of users we selected is 151,468. For the time limits reason, we downloaded all pictures of the selection of 27,538 users. Based on these data, we addressed the task of identification of professional photographers and non-professional users on Flickr. We used several feature sets and tested four models on them and their representation. From interpretable classification techniques -- Gaussian Naïve Bayes and LR and non-interpretable techniques -- RF and Gradient Boosting Classifier, RF showed the best performance using user and photo features. Our results demonstrated that it is feasible to properly predict whether a user is a professional photographer or not based on self-reported occupation labels. We also deduced that the technical and aesthetic scores of the picture are not highly correlated with the social activity carried out in this picture. Finally, based on statistically significant features, we draw the inference that professional photographers can be distinguished from non-professional users by higher NIMA aesthetic score, the average NIMA technical score, the average Kong score, the average number of groups, the average number of views and the number of the following users.

We will devote our future work to the models generalisation to detect professional photographers in other photo and video platforms that we described in this study as sites holding the potential for this type of task. Moreover, we will be expanding and replicating the study in other environments. Also, there is an assured need of performing the validation of our findings, e.g., through manual labelling or through other platforms such as LinkedIn. Besides, following the presented results, we would like to explore additional potential applications of this work, e.g., automatic detection of good photographers on the Web.


\bibliographystyle{unsrtnat}  
\bibliography{references}  

\end{document}

%% file: tables/platforms_comparison.tex
\begin{table*}[]
\resizebox{\textwidth}{!}{%
\begin{tabular}{|l|l|l|l|c|l|l|c|c|c|c|}
\hline
\begin{tabular}[c]{@{}l@{}}Photo \&\\ Video portal\end{tabular} & \begin{tabular}[c]{@{}l@{}}Foundation\\ year\end{tabular} & \begin{tabular}[c]{@{}l@{}}Area\\ served\end{tabular} & \begin{tabular}[c]{@{}l@{}}Languages\\ available\end{tabular} & \begin{tabular}[c]{@{}c@{}}Registration to\\ browse/contribute\end{tabular} & Free version & \begin{tabular}[c]{@{}l@{}}Number of\\ monthly users\end{tabular} & API & \begin{tabular}[c]{@{}c@{}}Photo editing\\ features\end{tabular} & \begin{tabular}[c]{@{}c@{}}Community of\\ photographers\end{tabular} & \begin{tabular}[c]{@{}c@{}}Ability to\\ write comments\end{tabular} \\ \hline
Flickr & 2004 & Worldwide & \begin{tabular}[c]{@{}l@{}}10 languages\\ incl. English\end{tabular} & \notick/\yestick & 1,000 photos & \begin{tabular}[c]{@{}l@{}}Over 60\\ million\end{tabular} & \yestick & \yestick & \yestick & \yestick \\ \hline
500px & 2009 & Worldwide & English & \notick/\yestick & \begin{tabular}[c]{@{}l@{}}7 photos per week\\ 2000 photos\end{tabular} & N/A & \notick & \notick & \yestick & \yestick \\ \hline
Instagram & 2010 & Worldwide & \begin{tabular}[c]{@{}l@{}}32 languages\\ incl. English\end{tabular} & Limited/\yestick & Unlimited photos & \begin{tabular}[c]{@{}l@{}}Over 2,000\\ million\end{tabular} & \yestick & \yestick & \yestick & \yestick \\ \hline
1x.com & 2007 & Worldwide & English & \notick/\yestick & \multicolumn{1}{c|}{\notick} & N/A & \notick & \notick & \yestick & \yestick \\ \hline
SmugMug & 2002 & Worldwide & English & \notick/\yestick & Up to 500MB & N/A & \yestick & \yestick & \yestick & \yestick \\ \hline
Pinterest & 2010 & Worldwide & \begin{tabular}[c]{@{}l@{}}37 languages\\ incl. English\end{tabular} & \notick/\yestick & 200,000 photos & \begin{tabular}[c]{@{}l@{}}Over 430\\ million\end{tabular} & \yestick & \notick & \yestick & \yestick \\ \hline
\end{tabular}%
}
\caption{Photo \& Video Sharing Platforms Comparison}
\label{tab:platforms_comparison}
\end{table*}

%% file: tables/results_models.tex
\begin{table}[]
\centering
\resizebox{0.495\textwidth}{!}{%
\begin{tabular}{|cllll|}
\hline
\multicolumn{1}{|c|}{Algorithm} & \multicolumn{1}{l|}{Feature set} & \multicolumn{1}{l|}{Accuracy} & \multicolumn{1}{l|}{AUC} & {F1 score} \\ \hline
\multicolumn{1}{|c|}{\multirow{7}{*}{\begin{tabular}[c]{@{}c@{}}Gaussian\\ Naïve \\ Bayes\end{tabular}}} & \multicolumn{1}{l|}{\begin{tabular}[c]{@{}l@{}}Crowdsourced\\ features\end{tabular}} & \multicolumn{1}{l|}{0.86} & \multicolumn{1}{l|}{0.51} & 0.85 \\ \cline{2-5} 
\multicolumn{1}{|c|}{} & \multicolumn{1}{l|}{User features} & \multicolumn{1}{l|}{0.86} & \multicolumn{1}{l|}{0.53} & 0.85 \\ \cline{2-5} 
\multicolumn{1}{|c|}{} & \multicolumn{1}{l|}{Photo features} & \multicolumn{1}{l|}{0.85} & \multicolumn{1}{l|}{0.52} & 0.85 \\ \cline{2-5} 
\multicolumn{1}{|c|}{} & \multicolumn{1}{l|}{\begin{tabular}[c]{@{}l@{}}Crowdsourced + \\ user features\end{tabular}} & \multicolumn{1}{l|}{0.92} & \multicolumn{1}{l|}{0.5} & 0.85 \\ \cline{2-5} 
\multicolumn{1}{|c|}{} & \multicolumn{1}{l|}{\begin{tabular}[c]{@{}l@{}}Crowdsourced + \\ photo features\end{tabular}} & \multicolumn{1}{l|}{0.85} & \multicolumn{1}{l|}{0.52} & 0.86 \\ \cline{2-5} 
\multicolumn{1}{|c|}{} & \multicolumn{1}{l|}{\begin{tabular}[c]{@{}l@{}}User + \\ photo features\end{tabular}} & \multicolumn{1}{l|}{0.86} & \multicolumn{1}{l|}{0.53} & 0.85 \\ \cline{2-5} 
\multicolumn{1}{|c|}{} & \multicolumn{1}{l|}{All features} & \multicolumn{1}{l|}{0.92} & \multicolumn{1}{l|}{0.5} & 0.85 \\ \hline
\multicolumn{5}{|l|}{} \\ \hline
\multicolumn{1}{|c|}{\multirow{7}{*}{\begin{tabular}[c]{@{}c@{}}Logistic \\ Regression\end{tabular}}} & \multicolumn{1}{l|}{\begin{tabular}[c]{@{}l@{}}Crowdsourced\\ features\end{tabular}} & \multicolumn{1}{l|}{0.92} & \multicolumn{1}{l|}{0.57} & 0.89 \\ \cline{2-5} 
\multicolumn{1}{|c|}{} & \multicolumn{1}{l|}{User features} & \multicolumn{1}{l|}{0.92} & \multicolumn{1}{l|}{0.66} & 0.89 \\ \cline{2-5} 
\multicolumn{1}{|c|}{} & \multicolumn{1}{l|}{Photo features} & \multicolumn{1}{l|}{0.92} & \multicolumn{1}{l|}{0.68} & 0.88 \\ \cline{2-5} 
\multicolumn{1}{|c|}{} & \multicolumn{1}{l|}{\begin{tabular}[c]{@{}l@{}}Crowdsourced + \\ user features\end{tabular}} & \multicolumn{1}{l|}{0.92} & \multicolumn{1}{l|}{0.53} & 0.88 \\ \cline{2-5} 
\multicolumn{1}{|c|}{} & \multicolumn{1}{l|}{\begin{tabular}[c]{@{}l@{}}Crowdsourced + \\ photo features\end{tabular}} & \multicolumn{1}{l|}{0.92} & \multicolumn{1}{l|}{0.67} & 0.89 \\ \cline{2-5} 
\multicolumn{1}{|c|}{} & \multicolumn{1}{l|}{\begin{tabular}[c]{@{}l@{}}User + \\ photo features\end{tabular}} & \multicolumn{1}{l|}{0.92} & \multicolumn{1}{l|}{0.63} & 0.88 \\ \cline{2-5} 
\multicolumn{1}{|c|}{} & \multicolumn{1}{l|}{All features} & \multicolumn{1}{l|}{0.92} & \multicolumn{1}{l|}{0.61} & 0.88 \\ \hline
\multicolumn{5}{|l|}{} \\ \hline
\multicolumn{1}{|c|}{\multirow{7}{*}{\begin{tabular}[c]{@{}c@{}}Random \\ Forest\end{tabular}}} & \multicolumn{1}{l|}{\begin{tabular}[c]{@{}l@{}}Crowdsourced\\ features\end{tabular}} & \multicolumn{1}{l|}{0.92} & \multicolumn{1}{l|}{0.58} & 0.88 \\ \cline{2-5} 
\multicolumn{1}{|c|}{} & \multicolumn{1}{l|}{User features} & \multicolumn{1}{l|}{0.92} & \multicolumn{1}{l|}{0.69} & 0.88 \\ \cline{2-5} 
\multicolumn{1}{|c|}{} & \multicolumn{1}{l|}{Photo features} & \multicolumn{1}{l|}{0.92} & \multicolumn{1}{l|}{0.68} & 0.88 \\ \cline{2-5} 
\multicolumn{1}{|c|}{} & \multicolumn{1}{l|}{\begin{tabular}[c]{@{}l@{}}Crowdsourced + \\ user features\end{tabular}} & \multicolumn{1}{l|}{0.92} & \multicolumn{1}{l|}{0.53} & 0.88 \\ \cline{2-5} 
\multicolumn{1}{|c|}{} & \multicolumn{1}{l|}{\begin{tabular}[c]{@{}l@{}}Crowdsourced + \\ photo features\end{tabular}} & \multicolumn{1}{l|}{0.92} & \multicolumn{1}{l|}{0.64} & 0.88 \\ \cline{2-5} 
\multicolumn{1}{|c|}{} & \multicolumn{1}{l|}{\begin{tabular}[c]{@{}l@{}}User + \\ photo features\end{tabular}} & \multicolumn{1}{l|}{0.92} & \multicolumn{1}{l|}{0.76} & 0.89 \\ \cline{2-5} 
\multicolumn{1}{|c|}{} & \multicolumn{1}{l|}{All features} & \multicolumn{1}{l|}{0.92} & \multicolumn{1}{l|}{0.65} & 0.88 \\ \hline
\multicolumn{5}{|l|}{} \\ \hline
\multicolumn{1}{|c|}{\multirow{7}{*}{\begin{tabular}[c]{@{}c@{}}Gradient\\ Boosting\\ Classifier\end{tabular}}} & \multicolumn{1}{l|}{\begin{tabular}[c]{@{}l@{}}Crowdsourced\\ features\end{tabular}} & \multicolumn{1}{l|}{0.92} & \multicolumn{1}{l|}{0.61} & 0.89 \\ \cline{2-5} 
\multicolumn{1}{|c|}{} & \multicolumn{1}{l|}{User features} & \multicolumn{1}{l|}{0.92} & \multicolumn{1}{l|}{0.72} & 0.88 \\ \cline{2-5} 
\multicolumn{1}{|c|}{} & \multicolumn{1}{l|}{Photo features} & \multicolumn{1}{l|}{0.92} & \multicolumn{1}{l|}{0.7} & 0.88 \\ \cline{2-5} 
\multicolumn{1}{|c|}{} & \multicolumn{1}{l|}{\begin{tabular}[c]{@{}l@{}}Crowdsourced + \\ user features\end{tabular}} & \multicolumn{1}{l|}{0.92} & \multicolumn{1}{l|}{0.57} & 0.89 \\ \cline{2-5} 
\multicolumn{1}{|c|}{} & \multicolumn{1}{l|}{\begin{tabular}[c]{@{}l@{}}Crowdsourced + \\ photo features\end{tabular}} & \multicolumn{1}{l|}{0.92} & \multicolumn{1}{l|}{0.68} & 0.88 \\ \cline{2-5} 
\multicolumn{1}{|c|}{} & \multicolumn{1}{l|}{\begin{tabular}[c]{@{}l@{}}User + \\ photo features\end{tabular}} & \multicolumn{1}{l|}{0.92} & \multicolumn{1}{l|}{0.69} & 0.88 \\ \cline{2-5} 
\multicolumn{1}{|c|}{} & \multicolumn{1}{l|}{All features} & \multicolumn{1}{l|}{0.92} & \multicolumn{1}{l|}{0.66} & 0.88 \\ \hline
\end{tabular}}%
\caption{Results comparison of Gaussian Naïve Bayes, Logistic Regression (LR), Random Forest and Gradient Boosting Classifier models by accuracy, AUC and F1 score metrics in the set of features -- crowdsourced, user, photo and their combinations.}
\label{tab:results_models}
\end{table}

%% file: tables/rq2_results.tex
\begin{table}[]
\resizebox{\columnwidth}{!}{%
\begin{tabular}{|cllll|}
\hline
\multicolumn{1}{|c|}{Algorithm} & \multicolumn{1}{l|}{Feature set} & \multicolumn{1}{l|}{Accuracy} & \multicolumn{1}{l|}{AUC} & {F1 score} \\ \hline
\multicolumn{1}{|c|}{\begin{tabular}[c]{@{}l@{}}Random\\ Forest\end{tabular}} & \multicolumn{1}{l|}{\begin{tabular}[c]{@{}l@{}}Aesthetics and \\technical features\end{tabular}} & \multicolumn{1}{l|}{0.92} & \multicolumn{1}{l|}{0.67} & 0.88 \\ \cline{2-5} 
\multicolumn{1}{|c|}{} & \multicolumn{1}{l|}{\begin{tabular}[c]{@{}l@{}}Social activity\\ features\end{tabular}}  & \multicolumn{1}{l|}{0.92} & \multicolumn{1}{l|}{0.6} & 0.88 \\ \hline
\end{tabular}%
}
\caption{Results comparison of RF model by accuracy, AUC and F1 score metrics in the aesthetics and technical quality features and social activity features.}
\label{tab:rq2}
\end{table}